\documentclass{nature}

\usepackage{amsmath}
\usepackage[pdftex]{graphicx}
\usepackage{dcolumn}  
\usepackage{bm}           
\usepackage{dsfont}
\usepackage{color}
\usepackage{braket}

\newcounter{defcounter}
\usepackage{amsthm}

\title{Quantum Vibronic Dynamics Shape Catalytically Relevant Au-Ligand Interfaces in Atomically Precise Gold Nanoclusters} 

\author{Mengyuan Cui$^{1,*}$,Tianrui Chen$^{1,*}$, Junhua Zhou$^{1,*}$, Xiangmei Duan$^{1}$, Vandana Tiwari$^{2,3}$, Chao Mei$^{1}$, Ajay Jha$^{4,5}$, Fulu Zheng$^{1}$, Hong-Guang Duan$^{1}$ } 

\begin{document} 

\maketitle 

\begin{affiliations} 
\item Department of Physics, School of Physical Science and Technology, Ningbo University, Ningbo, 315211, P.R. China 
\item Stanford PULSE Institute, SLAC National Accelerator Laboratory, Menlo Park, CA 94025
\item Department of Chemical Science, Linac Coherent Light Source, SLAC National Accelerator Laboratory, Menlo Park, CA 94025
\item Rosalind Franklin Institute, Harwell, Oxfordshire OX11 0QX, United Kingdom 
\item Department of Pharmacology, University of Oxford, Oxford, OX1 3QT United Kingdom  \\ 
\centerline{\underline{\date{\bf \today}}} 
\end{affiliations} 

\begin{abstract} 

Atomically precise gold nanoclusters have emerged as versatile platforms for photocatalysis, energy conversion and molecular photochemistry because their electronic structure is governed by strongly coupled metal-ligand interactions. Yet these interactions are almost exclusively discussed in terms of static atomic structures, leaving unresolved how photoexcitation dynamically reorganizes catalytically relevant Au-ligand interfaces before thermal equilibrium is established. Here, we investigate the rod-shaped nanocluster [Au$_{25}$(PPh$_3$)$_{10}$(SC$_2$H$_5$)$_5$Cl$_2$]$^{2+}$ using ultrafast transient-grating spectroscopy, two-dimensional electronic spectroscopy (2DES), {\em ab-initio} electronic-structure calculations and Hierarchical Equations of Motion (HEOM) simulations. The multidimensional spectra resolve multiple electronic relaxation pathways together with a hierarchy of coherent structural motions extending from localized Au-ligand distortions to collective vibrations of the gold framework. Wavelet analysis demonstrates that high-frequency Au-ligand vibrations are generated immediately following photoexcitation, whereas low-frequency collective modes emerge subsequently through interstate vibronic coupling, revealing sequential redistribution of structural coherence throughout the nanocluster. Theoretical simulations reproduce the nonlinear spectroscopic response and identify the microscopic vibronic couplings responsible for this hierarchical energy flow. Together, the results show that photoexcitation drives continuous ultrafast reorganization of the Au-ligand bonding network rather than simple electronic relaxation, transiently reshaping the electronic structure of catalytically relevant interfacial motifs before thermalization. More broadly, this work establishes dynamic Au-ligand interfaces as the microscopic link between excited-state energy flow and photochemical function in atomically precise metal nanoclusters.

\end{abstract} 


Catalysis at noble-metal active sites is governed not only by their static atomic structure, but also by the manner in which absorbed energy is redistributed among electronic, vibrational and interfacial degrees of freedom before thermal equilibrium is established \cite{Ref1, Ref2}. This issue becomes particularly acute in ultrasmall metal clusters \cite{Ref3, Ref4, Ref5}, where the absorption of a single photon can either generate chemically productive electronic reorganization or be dissipated through ultrafast non-radiative relaxation \cite{Ref6}. Atomically precise gold nanoclusters provide a uniquely powerful platform for addressing this problem because they occupy the boundary between molecular and metallic matter \cite{Ref7, Ref8, Ref9, Ref10}. As the cluster size approaches the quantum-confined regime, collective plasmonic behavior disappears and discrete electronic transitions emerge \cite{Ref11, Ref12}. Consequently, subtle changes in composition, ligand environment or geometric structure can profoundly alter charge localization, excited-state relaxation and energy-conversion pathways \cite{Ref13, Ref14}. These characteristics have established gold nanoclusters as model systems for fundamental nanochemistry and as promising materials for catalysis, photonics, sensing and solar-energy conversion.

The prototypical Au$_{25}$ family has played a central role in developing this understanding \cite{Ref15, Ref16}. Single-crystal structural studies revealed that Au$_{25}$(SR)$_{18}$- consists of an Au$_{13}$ core protected by metal-thiolate staple motifs, providing one of the first atomically resolved examples of a quantum-confined metal cluster \cite{Ref17}. Electronic-structure analyses subsequently demonstrated that the frontier states possess superatomic character, whereas deeper occupied and higher-lying virtual manifolds contain substantial contributions from Au 5d orbitals, sulphur 3p states and ligand-derived orbitals \cite{Ref16, Ref18}. Therefore, Au$_{25}$ cannot be viewed simply as either a metallic nanoparticle or a conventional coordination complex. Rather, it represents a hybrid metal-ligand electronic system in which frontier-orbital symmetry, ligand-field interactions and metal-ligand covalency collectively determine its optical and chemical properties. This intricate electronic structure suggests that excited-state dynamics in Au$_{25}$ should be governed not solely by electronic relaxation, but by a complex interplay between electronic and nuclear degrees of freedom.

Ultrafast studies sharpened that picture, but they also exposed its limits. Two-dimensional electronic spectroscopy on Au$_{25}$(SC$_8$H$_9$)$_{18}$- resolved hot-electron relaxation within the superatom D manifold on a $\sim$200 fs time scale and hot-hole relaxation within the P manifold on a $\sim$300 fs time scale, proving that excitation-detection correlations can isolate pathways obscured in one-dimensional transient absorption \cite{Ref19}. Complementary transient absorption and photoluminescence work then showed that interband relaxation can depend strongly on ligand identity and solvent dielectric through charge-transfer-assisted channels, whereas intraband relaxation within core-localized superatom states is much less sensitive to the environment and can be mediated by low-frequency Au-Au motion \cite{Ref20}. Another femtosecond study of [Au$_{25}$(SR)$_{18}$]- revealed coherent oscillations near 40 and 80 cm$^{-1}$, assigned to acoustic motion of the Au$_{13}$ core, with phase behavior consistent with amplitude modulation for one mode and frequency modulation for the other \cite{Ref21}. More recent work on Au$_{38}$ extended the same logic by showing that excited-state symmetry can redirect relaxation pathways in dense nanocluster manifolds \cite{Ref22}. Together, these studies have established a detailed picture of the electronic states involved in relaxation and have identified key low-frequency vibrational modes. Yet they leave unanswered how these electronic and vibrational degrees of freedom interact dynamically to govern energy flow following photoexcitation.

This unresolved issue represents a deeper and more fundamental problem. Previous investigations have largely addressed either which electronic states relax or which vibrational modes are excited, but they have not established how an initially prepared electronic wavepacket generates, transfers and redistributes vibronic coherence. Such a distinction is crucial because catalytic function ultimately depends on how nuclear motion reorganizes electronic structure before thermalization occurs \cite{Ref23}. The central question therefore is whether low-frequency vibrations merely act as passive heat sinks that broaden spectral lines and dissipate excess energy, or whether they actively mediate population transfer, coherence exchange and electronic-state mixing through vibronic coupling. In quantum-confined nanoclusters, where relevant electronic energy gaps are often comparable to characteristic vibrational frequencies, this distinction cannot be regarded as a perturbative correction to electronic kinetics. Rather, vibronic interactions may constitute the dominant mechanism by which electronic excitation is transformed into chemically relevant nuclear motion. Understanding how this conversion occurs, how coherence propagates through the metal-ligand framework, and how such dynamics reshape the excited-state electronic landscape represents one of the outstanding challenges in the photochemistry of atomically precise nanoclusters.

The importance of this problem is amplified in the rod-shaped [Au$_{25}$(PPh$_3$)$_{10}$(SC$_2$H$_5$)$_5$Cl$_2$]$^{2+}$ complex (Au$_{25}$-rod). In contrast to the highly symmetric thiolate-protected Au$_{25}$(SR)$_{18}$- superatom, the simultaneous presence of phosphine, thiolate and chloride ligands generates a markedly anisotropic ligand field and introduces stronger mixing between metal-centred and ligand-centred electronic and vibrational degrees of freedom. Such complexity creates an ideal platform for investigating how vibronic interactions emerge within a chemically heterogeneous metal cluster. Resolving this problem requires direct access to correlated electronic and vibrational dynamics with state specificity. Two-dimensional electronic spectroscopy \cite{Ref24, Ref25, Ref26} is uniquely suited to this challenge because phase-matched measurements isolate the third-order nonlinear response and project overlapping relaxation pathways into excitation-detection frequency space, enabling the temporal evolution of individual diagonal and cross-peaks to be followed independently \cite{Ref27, Ref28, Ref29, Ref30}. Here, we combine ultrafast 2DES measurements with {\em ab initio} excited-state electronic-structure calculations, phase-matched simulations of the nonlinear optical response, and the quantum-dynamical modeling to establish a microscopic description of excited-state energy flow in Au$_{25}$-rod. This combined approach reveals distinct electronic couplings, multiple relaxation timescales and a hierarchy of coherent motions spanning from tens to more than one thousand wavenumbers, providing a framework for understanding how vibronic interactions govern the conversion of electronic excitation into chemically relevant nuclear dynamics in atomically precise metal nanoclusters.

\section*{Results} 

\begin{figure}[h!]
\begin{center}
\includegraphics[width=16.0cm]{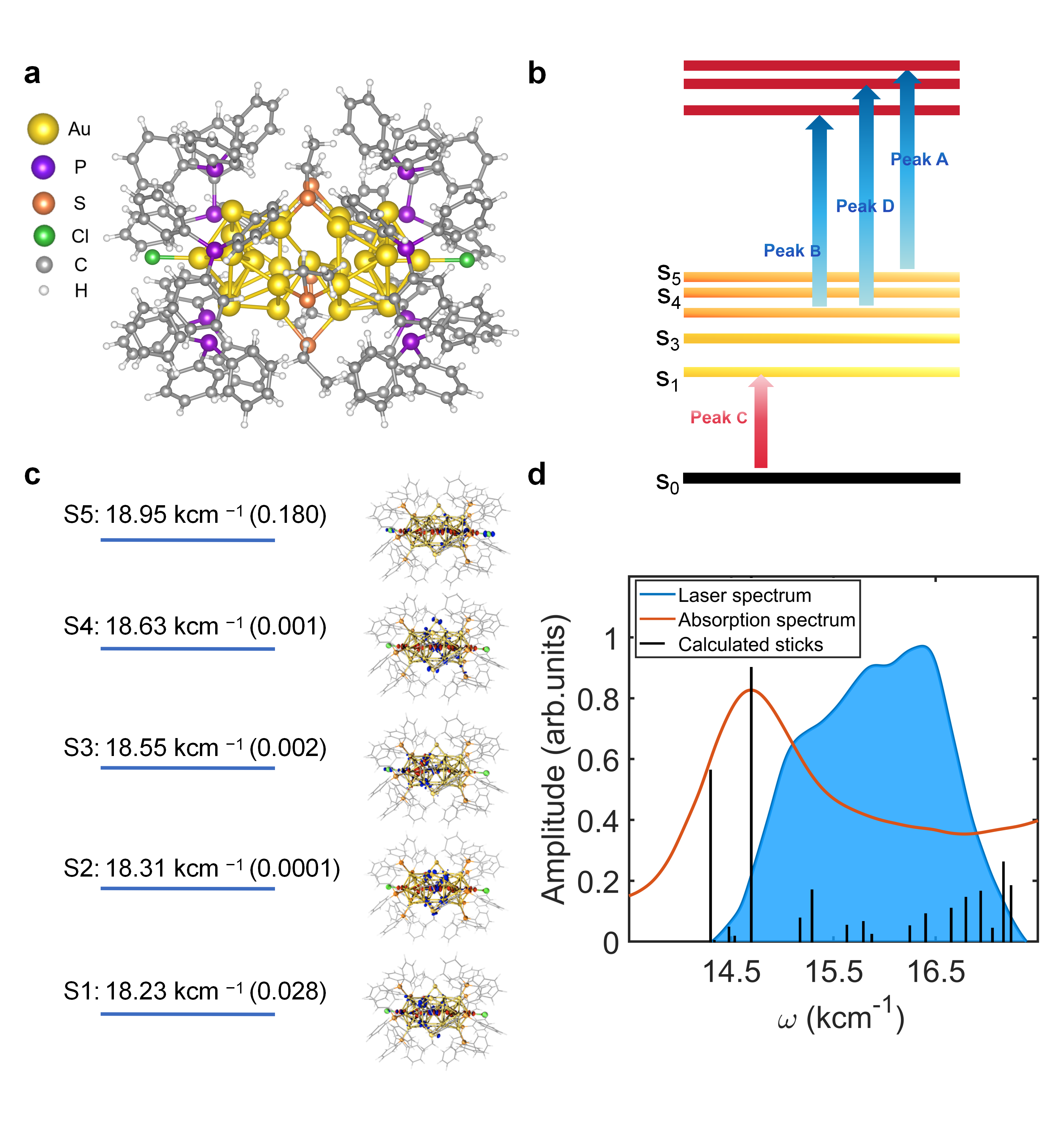}
\caption{\label{fig:Fig1}  Electronic structure of the dynamic Au-ligand interface. (a) Molecular structure of the atomically precise Au$_{25}$-rod nanocluster. (b) Schematic representation of the electronic ground state, selected excited states and higher-lying excited states involved in the optical response. (c) Calculated energies of the low-lying excited states together with their dominant molecular-orbital contributions. (d) Experimental absorption spectrum of Au$_{25}$-rod (red) and excitation spectrum used in the measurements (blue shaded region). Calculated low-lying excited states and their corresponding transition-dipole strengths are also indicated.}
\end{center}
\end{figure}

The molecular structure of the atomically precise rod-shaped gold nanocluster Au$_{25}$-rod is shown in Fig.\ \ref{fig:Fig1}(a). The cluster consists of a gold framework protected by phosphine, thiolate and chloride ligands, creating an anisotropic metal-ligand environment that gives rise to a rich manifold of electronic and vibrational states. The schematic representation of electronic ground (S$_{0}$), excited (S$_{1}$, S$_{3}$, S$_{4}$, S$_{5}$) and higher excited states (denoted by deep red bars) is shown in Fig.\ \ref{fig:Fig1}(b). The low-lying triplet manifold was also examined to assess whether spin-state conversion contributes to the observed dynamics. As detailed in Section X of the Supporting Information, the lowest triplet states are energetically separated from the corresponding singlet states and the calculated spin-orbit couplings between S$_{1}$ and the low-lying triplet manifold are only a few cm$^{-1}$. Together with the absence of an experimentally resolved kinetic component that can be specifically assigned to triplet formation, these calculations indicate that the early-time dynamics investigated here are dominated by relaxation within the singlet manifold. We therefore focus the subsequent vibronic analysis and HEOM simulations on the singlet excited-state landscape. 

The optical transitions of selected peaks in 2DES have been labeled as peaks A to D in Fig.\ \ref{fig:Fig1}(b). The experimentally measured absorption spectrum is presented in Fig.\ \ref{fig:Fig1}(d) together with the excitation spectrum employed in this work. The broad overlap between the two spectra enables selective access to the low-energy electronic transitions of the nanocluster. To establish the underlying excited-state landscape, electronic-structure calculations were performed, and the resulting excited-state energies and orbital characters are summarized in Fig.\ \ref{fig:Fig1}(c). The calculations reveal multiple closely spaced excited states with mixed metal- and ligand-derived character within the spectral region addressed by the excitation pulse. The combination of a structurally well-defined metal cluster, spectrally accessible excited states and a dense manifold of electronic transitions make this system particularly suitable for investigating ultrafast excited-state dynamics. 

\subsection{Resolving excited-state evolution with 2DES.} 

\begin{figure}[h!]
\begin{center}
\includegraphics[width=15.0cm]{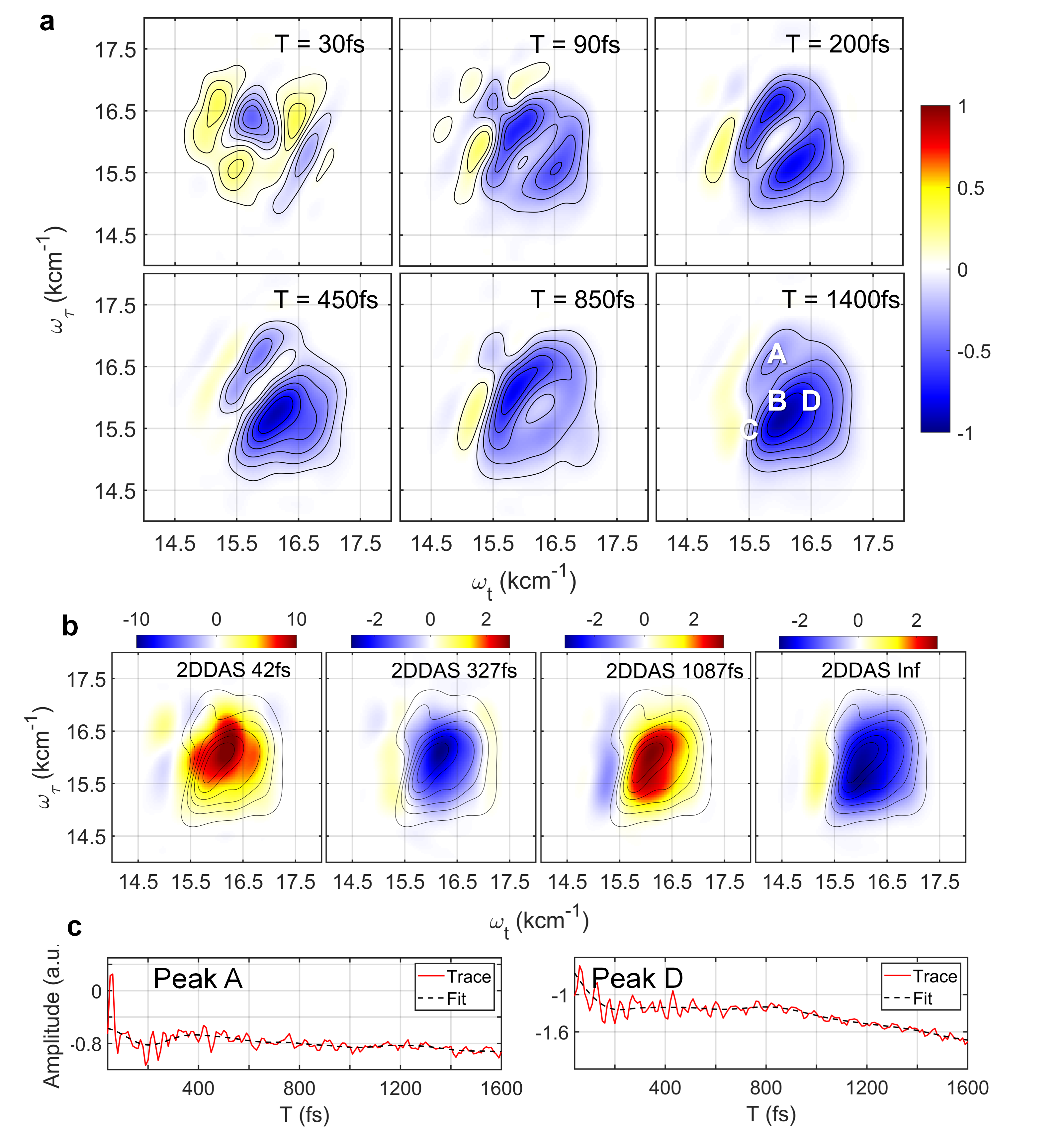}
\caption{\label{fig:Fig2}  Ultrafast reorganisation of the excited-state manifold. (a) Experimental absorptive 2DES spectra of Au$_{25}$-rod at representative waiting times of 30, 90, 200, 450, 850 and 1400 fs. The principal spectral features analysed in this work are labelled A-D. (b) Two-dimensional decay-associated spectra (2D-DAS) corresponding to characteristic timescales of 42 fs, 327 fs, 1087 fs and a long-lived component. Contours of the experimental spectrum at 1.4 ps are overlaid for comparison. (c) Waiting-time traces extracted from representative peaks A and D, illustrating the multiscale evolution of the excited-state response. }
\end{center}
\end{figure}

The real part of the measured 2DES is shown in Fig.\ \ref{fig:Fig2}(a) for representative waiting times of 30, 90, 200, 450, 850 fs and 1.4 ps. Positive and negative amplitudes correspond to ground-state bleach/stimulated emission (GSB/SE) and excited-state absorption (ESA), respectively. Immediately following photo-excitation, the spectra exhibit substantial temporal evolution. At T = 30 fs (Fig.\ \ref{fig:Fig2}a), positive and negative spectral contributions strongly overlap, producing a congested pattern of diagonal and off-diagonal features. By T = 90 fs in Fig.\ \ref{fig:Fig2}(a), the spectral distribution has changed substantially, with two distinct ESA features becoming visible. These features further evolve at T = 200 fs (Fig.\ \ref{fig:Fig2}a), where the ESA contributions become more pronounced. 

The spectral evolution continues at later waiting times. At T = 450 fs and 850 fs in Fig.\ \ref{fig:Fig2}(a), the relative amplitudes of the resolved features change while their spectral positions remain largely preserved. Even at T = 1.4 ps (Fig.\ \ref{fig:Fig2}(a)), multiple diagonal and cross-peaks remain clearly resolved. The excellent signal-to-noise ratio of the measurements allows individual spectral features to be tracked throughout the entire waiting-time window. For subsequent analysis, the principal peaks are labeled A-D in Fig.\ \ref{fig:Fig2}(a) at T = 1.4 ps. The persistence of well-resolved diagonal and cross-peaks throughout the measurement window demonstrates that the Au$_{25}$-rod cluster exhibits a rich and highly structured excited-state response that can be followed directly in excitation-detection frequency space. 

\subsection{Population dynamics revealed by peak-resolved kinetics. } 

To quantify the temporal evolution of the resolved spectral features, the amplitudes of peaks A and D were extracted from the 2DES and plotted as a function of waiting time, as shown in Fig.\ \ref{fig:Fig2}(c). All selected peaks display pronounced time-dependent amplitude variations across the measured time window. The corresponding fitting procedures and residuals are provided in the SI (Fig.\ S1 and Fig.\ S2). To identify characteristic kinetic components underlying the spectral evolution, a global analysis was performed on the complete three-dimensional 2DES dataset. The resulting two-dimensional decay-associated spectra (2DDAS) are shown in Fig.\ \ref{fig:Fig2}(b). Four principal components are resolved within the experimental time window. The fastest component, with a characteristic timescale of 42 fs, is shown in Fig.\ \ref{fig:Fig2}(b). A second component with a timescale of 327 fs is presented as well. A third component with a characteristic timescale of 1087 fs is shown in Fig.\ \ref{fig:Fig2}(b), while the long-lived contribution is presented as the last in Fig.\ \ref{fig:Fig2}(b). For comparison, contours of the experimental spectrum at T = 1.4 ps are overlaid on the corresponding 2DDAS maps. To extend the observation window beyond the femtosecond regime, additional 2DES measurements were performed up to 550 ps. Global analysis of the extended dataset revealed kinetic components with characteristic timescales of 86 fs, 329 fs, 1.7 ps, 15 ps, 43 ps and an effectively time-independent contribution. The corresponding spectra and analysis are provided in the SI (Fig.\ S3). Together, these results establish that the excited-state evolution of the Au$_{25}$-rod cluster proceeds through multiple distinct timescales spanning more than four orders of magnitude. 

\subsection{Observation of coherences across the excited-state manifold. }

\begin{figure}[h!]
\begin{center}
\includegraphics[width=16.0cm]{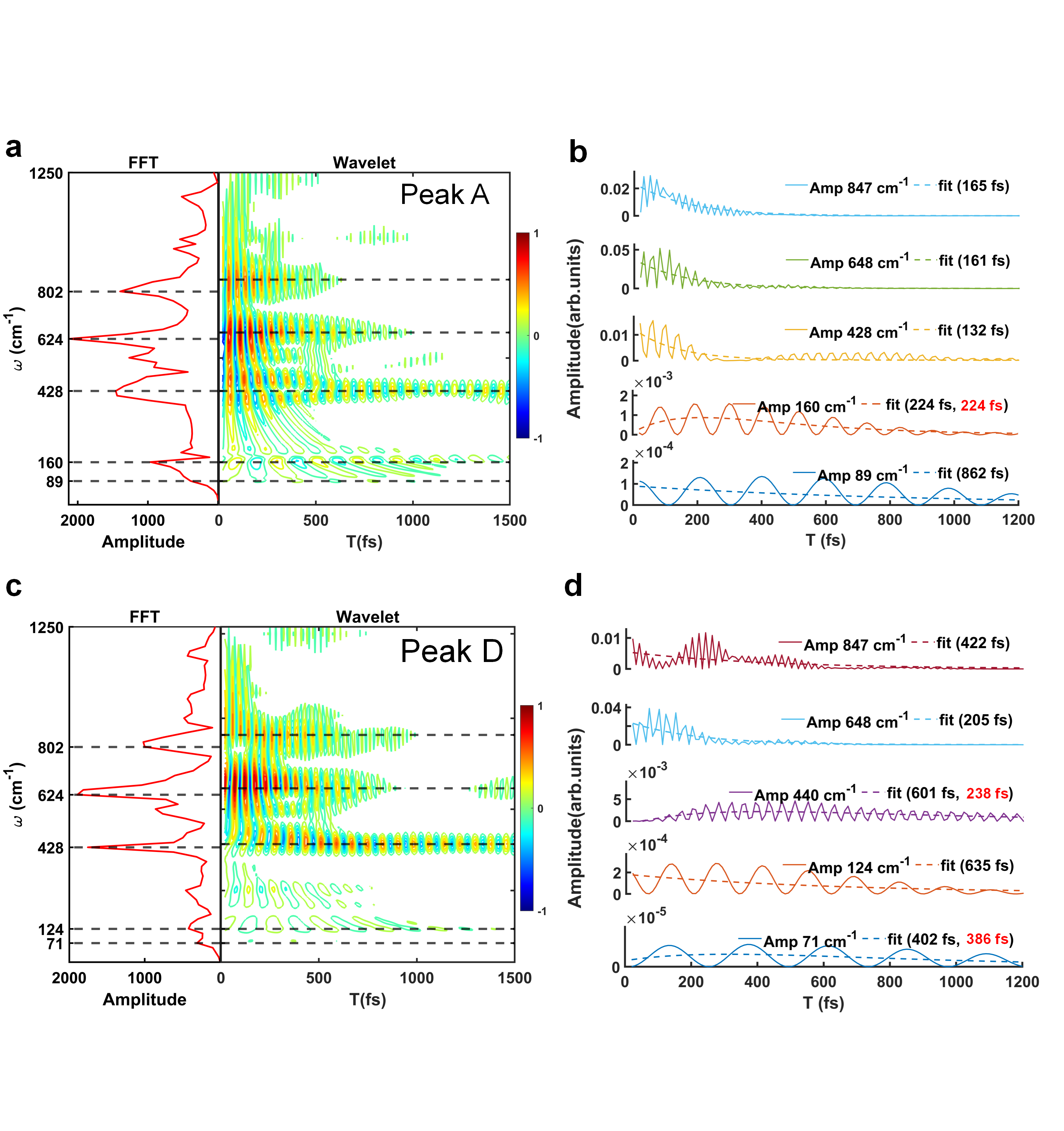} 
\caption{\label{fig:Fig3}  Hierarchical flow of vibronic coherence. (a,c) Fourier-transform spectra and corresponding wavelet analyses of the residual oscillatory signals extracted from peaks A and D after removal of the population kinetics. The spectra resolve multiple vibrational coherences spanning low-frequency collective motions and higher-frequency metal–ligand and ligand-centred vibrations. (b,d) Time-dependent amplitudes of the selected vibrational modes, revealing distinct coherence lifetimes and, for several modes, delayed growth prior to decay. The results demonstrate that coherent nuclear motion evolves differently across the excited-state manifold.} 
\end{center}
\end{figure}

To investigate coherent dynamics embedded within the 2DES response, global kinetic components were first removed from the waiting-time traces, and the residual oscillatory signals were analyzed using Fourier-transform and wavelet approaches. Representative results for peaks A and D are shown in Fig.\ \ref{fig:Fig3}. The Fourier-transform spectrum of peak A (Fig.\ \ref{fig:Fig3}a) reveals a series of well-defined oscillatory components. Wavelet analysis further resolves the temporal evolution of these frequencies across the waiting-time dimension. Eight prominent vibrational frequencies are identified at approximately 89, 160, 428, 552, 624 and 802 cm$^{-1}$. The corresponding time-dependent amplitudes are shown in Fig.\ \ref{fig:Fig3}(b).

For peak D, analogous analysis is presented in Fig.\ \ref{fig:Fig3}(c and d). Multiple oscillatory components are observed at frequencies of approximately 71, 124, 428, 624 and 802 cm$^{-1}$. The wavelet analysis demonstrates that these oscillations evolve throughout the waiting-time window and exhibit distinct temporal profiles. The high quality of the residual oscillatory signals permits extraction of characteristic coherence lifetimes for individual vibrational modes. The resulting values are summarized in Fig.\ \ref{fig:Fig3}(b,d). Similar analyses for peaks B and C are provided in the SI (Fig.\ S4). These measurements demonstrate that coherent vibrational dynamics are distributed across multiple spectral regions of the 2DES and span a broad frequency range extending from low-frequency collective motions to higher-frequency ligand-associated vibrations. 

\subsection {Microscopic origin of vibronic energy flow. } 

\begin{figure}[h!]
\begin{center}
\includegraphics[width=15.0cm]{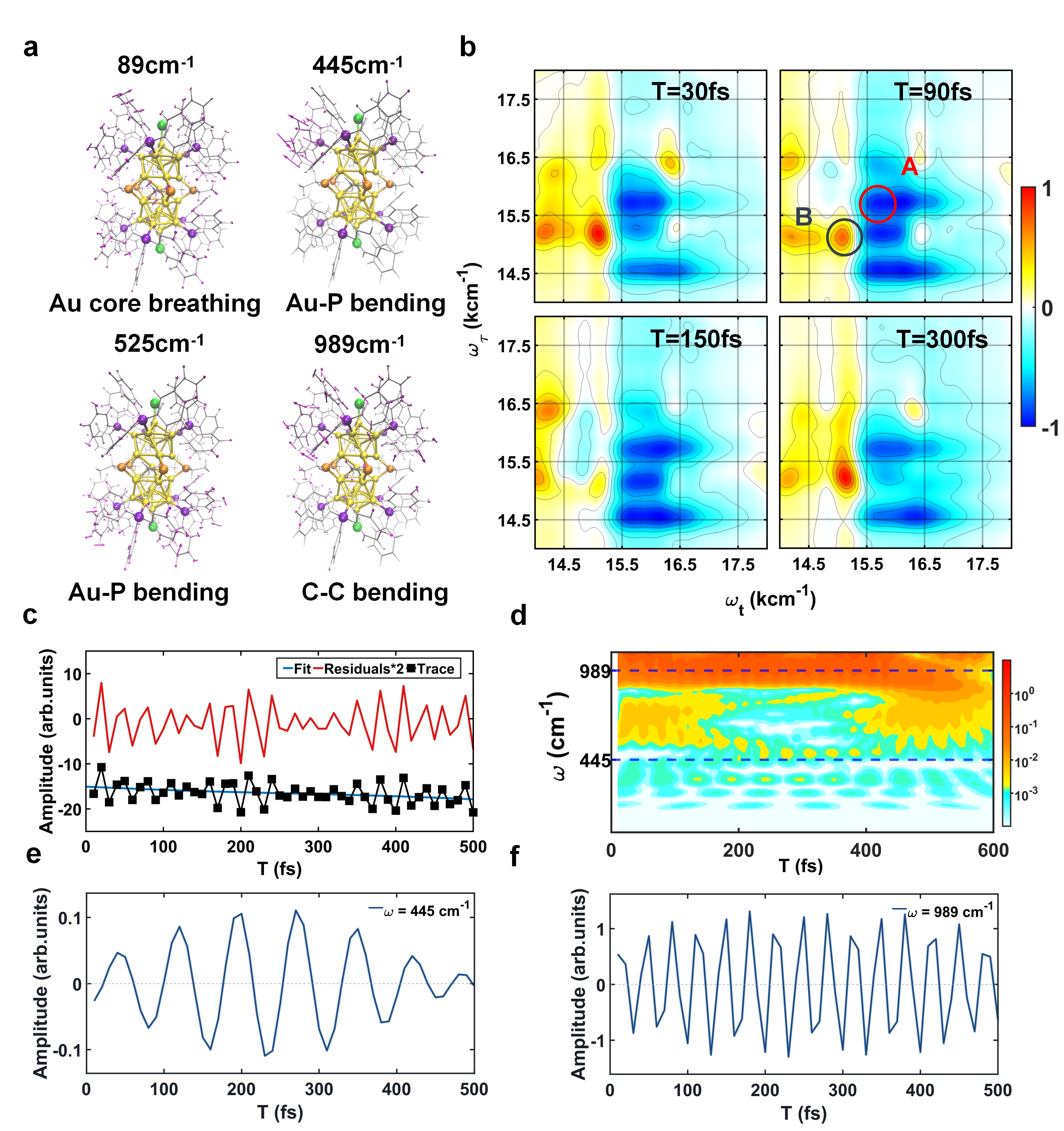} 
\caption{\label{fig:Fig4}  Microscopic origin of dynamic Au-ligand interfaces. (a) Representative calculated normal modes of the Au$_{25}$-rod nanocluster, illustrating structural motions ranging from collective Au-framework vibrations to localized Au-ligand distortions. (b) Simulated 2DES spectra at waiting times of 30, 90, 150 and 300 fs obtained from the vibronic HEOM model. (c) Calculated waiting-time trace of peak A together with the corresponding oscillatory residual. (d) Wavelet analysis of the residual signal. (e,f) Calculated coherent dynamics of the 445 and 989 cm$^{-1}$ modes, respectively, showing their distinct generation and decay timescales.} 
\end{center}
\end{figure}

To provide a microscopic interpretation of the experimentally observed dynamics, we performed electronic-structure calculations together with quantum-dynamical simulations of the nonlinear optical response. The optimized ground-state geometry reproduces the experimentally determined structure of the Au25-rod cluster and provides the basis for calculating the excited-state manifold shown in Fig.\ \ref{fig:Fig1}(c). The calculated vibrational modes are summarized in Fig.\ \ref{fig:Fig4}(a), where representative normal modes at 89, 445, 525 and 989 cm$^{-1}$ illustrate the hierarchy of structural motions spanning collective Au-framework vibrations and localized Au-ligand distortions. Additional vibrational frequencies, normal modes and vibronic coupling strengths are presented in Sections VI and VII of the SI. These calculations reveal that the optically accessible excited states involve mixed metal-centred and ligand-centred character, indicating that photoexcitation simultaneously perturbs both the Au framework and the surrounding ligand environment.

To establish how these electronic and vibrational degrees of freedom interact during excited-state evolution, the calculated electronic manifold was incorporated into an excitonic Hamiltonian propagated using the HEOM. The corresponding nonlinear optical response was subsequently calculated using the phase-matching formalism employed for simulating 2DES signals. Representative calculated spectra are presented in Fig.\ \ref{fig:Fig4}(b) for waiting times of 30, 90, 150 and 300 fs. The simulations reproduce the principal experimental observations, including the temporal evolution of excited-state absorption together with the appearance of distinct diagonal and cross-peaks, demonstrating that explicit treatment of vibronic coupling captures the essential excited-state dynamics.

To further examine the microscopic origin of the observed coherences, representative kinetics extracted from peak A are shown in Fig.\ \ref{fig:Fig4}(c), together with the corresponding residual oscillatory component. Wavelet analysis of these residuals (Fig. 4(d)) resolves two dominant vibrational contributions centred near 445 and 989 cm$^{-1}$. Their temporal evolution is presented in Fig. 4(e) and (f), respectively. The higher-frequency 989 cm$^{-1}$ mode develops within approximately the first 100 fs following photoexcitation, whereas the 445 cm$^{-1}$ vibration emerges on a slower timescale of approximately 200 fs. This sequential appearance indicates that coherent structural motion evolves hierarchically rather than being generated simultaneously through impulsive excitation. Similar analyses performed for additional vibrational modes are presented in Section IX of the SI.

\section*{Dynamic Au-Ligand Interfaces as the Microscopic Origin of Photochemical Function} 

\begin{figure}[h!]
\begin{center}
\includegraphics[width=13.0cm]{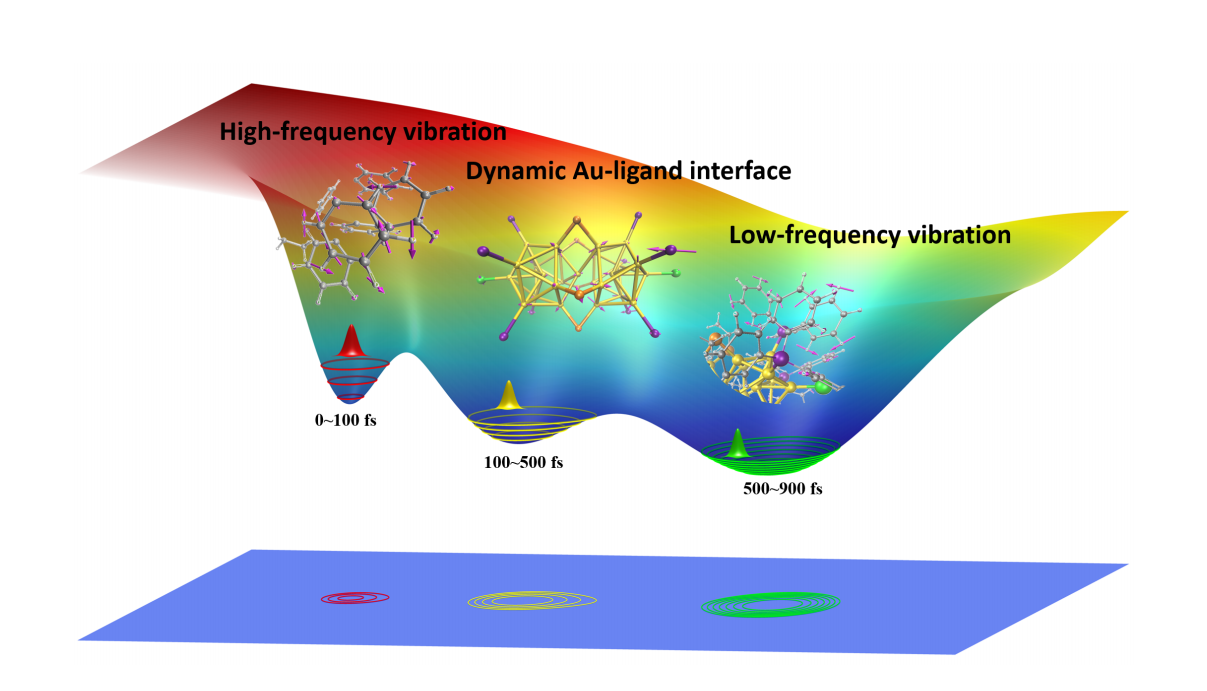} 
\caption{\label{fig:Fig5}  Dynamic Au-ligand interface generated through hierarchical vibronic energy flow. Schematic representation of the proposed excited-state mechanism. Photoexcitation initially generates localized high-frequency Au-ligand distortions, followed by interstate vibronic coupling and redistribution into lower-frequency collective motions of the metal-ligand framework. This hierarchical structural evolution dynamically reorganizes Au-Au and Au-ligand interactions, providing a microscopic connection between excited-state energy flow and catalytically relevant interfacial structure.  }
\end{center}
\end{figure}

The central question addressed in this work extends beyond identifying electronic relaxation times or cataloguing vibrational coherences. Instead, it concerns how photoexcitation transiently reshapes the atomic and electronic structure of an atomically precise nanocluster before thermal equilibrium is established. While previous ultrafast studies have identified hot-carrier relaxation and coherent Au-framework vibrations, the microscopic connection between excited-state energy flow and chemically relevant structural evolution has remained unresolved. The combined experimental and theoretical picture established here reveals that this evolution occurs predominantly within a vibronically coupled singlet manifold. Although low-lying triplet states are present, calculations show relatively weak spin-orbit coupling between S$_{1}$ and the lowest triplet states, while the experimental spectra reveal no kinetic or spectral component that can be unambiguously assigned to triplet formation. The early-time dynamics can therefore be described primarily in terms of redistribution within the singlet vibronic manifold, providing the basis for the mechanistic picture summarized in Fig.\ \ref{fig:Fig5}.

Following photoexcitation, the initially prepared electronic wavepacket drives local structural distortion of the Au-ligand framework. As summarized schematically in Fig.\ \ref{fig:Fig5}, high-frequency Au-ligand motions emerge within the first 100 fs, whereas lower-frequency collective motions develop subsequently on the few-hundred-femtosecond timescale. This temporal ordering, observed experimentally and reproduced by the vibronic simulations, indicates that the different coherent motions are not simply generated independently by the excitation pulse. Instead, interstate vibronic coupling provides a pathway through which initially localized structural motion is redistributed into increasingly collective coordinates involving the Au framework and ligand environment. This behaviour is reminiscent of coherence-transfer processes reported in lead-halide perovskites, where initially generated high-frequency motion evolves into lower-frequency structural distortion \cite{Ref31}.

This hierarchical redistribution fundamentally changes how coherent nuclear motion should be interpreted. Instead of viewing coherent oscillations simply as spectroscopic fingerprints of impulsive excitation, the present results demonstrate that coherent structural motion actively emerges during excited-state evolution itself. Vibronic coherence therefore becomes the physical mechanism through which electronic excitation reorganizes the structure of the nanocluster. Perhaps more importantly, these structural dynamics redefine the concept of catalytic active sites in atomically precise nanoclusters. Catalytic reactions are generally assumed to occur at exposed Au atoms or at Au-ligand interfacial motifs whose electronic structures are determined by equilibrium atomic geometries. Our results demonstrate that the Au-ligand framework cannot be regarded as structurally static immediately following photoexcitation. These motions are expected to dynamically modulate the electronic properties of catalytically relevant interfacial motifs.

The calculations provide direct microscopic support for this interpretation. High-frequency Au-ligand distortions initially localize structural motion around chemically distinct coordination environments, whereas lower-frequency collective modes subsequently redistribute these distortions throughout the gold framework. Such coordinated structural evolution is expected to transiently modify the electronic properties of Au-ligand interfacial motifs that are widely recognized as governing adsorption energies, charge-transfer efficiency and photocatalytic activity in atomically precise gold nanoclusters. Although the present work does not directly probe catalytic turnover, it identifies the fundamental excited-state structural dynamics that precede these chemical processes. This perspective naturally explains why relatively small changes in ligand identity often produce disproportionately large changes in photocatalytic performance. Ligands are not merely structural stabilizers but define the topology of excited-state potential-energy surfaces and therefore determine how absorbed optical energy propagates through the nanocluster. The efficiency of photochemical function is thus expected to depend not only on equilibrium electronic structure but also on the topology of vibronic couplings that governs ultrafast structural reorganization. More broadly, our results establish a conceptual bridge between multidimensional ultrafast spectroscopy and nanocluster photocatalysis. Rather than considering catalytic interfaces as static structural motifs, they emerge here as dynamically evolving electronic structures generated by coherent nuclear motion. This dynamic picture provides a microscopic framework through which optical excitation, structural reorganization and ultimately chemical reactivity become naturally connected.

\section*{Conclusion} 

The present work establishes a microscopic picture of how optical excitation is converted into structural dynamics in atomically precise gold nanoclusters. Combining multidimensional spectroscopy with electronic-structure calculations and quantum-dynamical simulations demonstrates that excited-state evolution proceeds through hierarchical vibronic interactions linking localized Au-ligand distortions with collective structural rearrangements of the gold framework. Rather than representing isolated vibrational signatures, these coherences define the structural pathways through which electronic excitation is redistributed throughout the nanocluster. More importantly, the results reveal that the Au-ligand interface itself evolves continuously during the first few hundred femtoseconds following photoexcitation. Such ultrafast structural reorganization transiently reshapes Au-Au bonding, metal-ligand covalency and electronic charge localization before thermal equilibrium is reached, indicating that catalytically relevant interfacial motifs should be regarded as dynamic rather than static entities under photoexcitation. Although the present study does not directly investigate catalytic reactions, it identifies the microscopic structural dynamics that precede and potentially regulate photocatalytic function. This dynamic description extends the conventional static picture of atomically precise nanoclusters and suggests that future strategies for designing photocatalysts may benefit from controlling not only equilibrium electronic structures but also the ultrafast vibronic pathways that govern transient structural evolution immediately following light absorption. 

%


\section*{Materials and Methods}
\subsection{Sample preparation.} 

The atomically precise gold nanocluster Au$_{25}$-rod was synthesized according to a previously reported protocol (Ref.\ \cite{sample1}) and used without further chemical modification. The purified complex was dissolved in dichloromethane and diluted to an optical density of approximately 0.2 at 620 nm. During spectroscopic measurements, the sample solution was continuously circulated through a 0.5 mm path length flow cell using a micropump system (Micropump GAX21-DEMSE) to minimize photodegradation and ensure a constant supply of fresh sample to the laser focus. To further optimize measurement conditions, the position of the excitation volume within the flow cell was adjusted using a computer-controlled three-axis translation stage. This procedure enabled maximization of signal quality while minimizing contributions from scattering and optical inhomogeneities. Additional details regarding laser characteristics and spectroscopic measurements are provided below. 

\subsection{Two-dimensional electronic spectroscopy and transient-grating measurements. } 

Ultrafast measurements were performed using a phase-stabilized all-reflective multidimensional spectrometer based on a diffractive-optics design \cite{Ref32, Ref33}. Broadband visible pulses were generated by a home-built non-collinear optical parametric amplifier (NOPA) pumped by a commercial femtosecond regenerative amplifier system. Pulse compression was achieved using a deformable mirror (OKO Technologies, 19 channels) in combination with a prism-pair compressor, yielding pulse duration of approximately 12 fs at the sample position. The temporal characteristics of the compressed pulses were verified by frequency-resolved optical gating (FROG), and the retrieved traces were analyzed using FROG3 software (Femtosoft Technologies). The resulting laser spectrum was centered at approximately 620 nm with a bandwidth of $\sim$100 nm (FWHM), providing simultaneous access to the low-energy electronic transitions of the Au$_{25}$ nanocluster. Three phase-controlled excitation pulses were focused onto the sample with a spot diameter of approximately 130 $\mu$m. The emitted third-order nonlinear signal was generated in the phase-matched direction and spectrally resolved using a Sciencetech 9055F imaging spectrograph coupled to a high-speed linear CCD detector.

Transient-grating measurements were acquired over a coherence-time window spanning -200 fs to 3 ps using 3 fs sampling intervals. At each delay position, 200 individual spectra were accumulated and averaged to improve the signal-to-noise ratio. The excitation pulse energy was maintained at approximately 10 nJ with a repetition rate of 1 kHz for all experiments. The phase of the transient-grating response was retrieved using the invariant-theorem procedure described previously \cite{Ref34}, enabling reconstruction of the absorptive signal used for subsequent multidimensional spectroscopic analysis \cite{Ref35}. 

\subsection{Theoretical calculations.} 

We performed full geometry optimization of Au$_{25}$ nanoclusters using the Gaussian 16 package at the BP86-D3(BJ)/def2-SVP level of theory. Based on the optimized structure, we calculated the normal vibrational modes and their Raman activities, with the Raman intensities obtained from polarizability derivatives, allowing direct comparison with experimental spectra. All computed frequencies are positive, confirming that the optimized structure corresponds to a stable minimum on the potential energy surface. Subsequently, to accurately describe the charge-transfer characteristics in the excited states, we employed the CAM-B3LYP long-range corrected hybrid functional together with the def2-SVP basis set to compute the excited-state properties of the cluster. For deeper insight into the nature of the excited-state transitions, we systematically analyzed the natural transition orbitals, electron–hole distributions, transition densities, and charge density differences. More details are described in Section VI of the Supporting Information.

For an open quantum system, the total Hamiltonian can be generally written as 
\begin{equation}
\label{eq:Htot}
H = H_S + H_B + H_{SB},
\end{equation}
where $H_S$ is the Hamiltonian of the system of interest, $H_B$ describes the surrounding bath, and $H_{SB}$ represents the system--bath interaction. In the present work, we consider a vibronic Hamiltonian that explicitly includes both electronic and vibrational degrees of freedom. The system consists of seven electronic states coupled to two intramolecular vibrational modes. The primary model discussed in the main text explicitly includes the 445 and 989 cm$^{-1}$ modes, selected to represent the experimentally observed hierarchy of structural dynamics. Alternative two-mode parameterizations were additionally examined as controls, as described in Section IX of the Supporting Information. The system Hamiltonian can be expressed as
\begin{equation}
\label{eq:Hs}
H_S = \sum_{s=0}^{6} |s\rangle \left( \epsilon_s + h_s \right)
\langle s| + \sum_{a<b} \left( |a\rangle V_{ab} \langle b| + |b\rangle V_{ab} \langle a| \right),
\end{equation}
where $|s\rangle$ denotes the $s$-th electronic state, $\epsilon_s$ is the corresponding electronic energy, and $h_s$ describes the vibrational Hamiltonian on the $s$-th electronic surface. The off-diagonal term $V_{ab}$ represents vibronic coupling between selected electronic states and enables population transfer and coherence between different electronic manifolds. The state-dependent vibrational Hamiltonian is written as 
\begin{equation}
\label{eq:hs}
h_s = \sum_{j=t,c} \Omega_j \left( a_j^\dagger a_j + \frac{1}{2} \right) + \sum_{j=t,c} \kappa_j^{(s)} Q_j,
\end{equation}
where $a_j^\dagger$ and $a_j$ are the bosonic creation and annihilation operators of the $j$-th vibrational mode, and $Q_j=a_j+a_j^\dagger$ is the corresponding vibrational coordinate. The parameter $\kappa_j^{(s)}$ describes the linear electron--vibrational coupling strength on electronic state $s$. This term modulates the electronic energy along the vibrational coordinates and allows electronic and vibrational degrees of freedom to mix explicitly. The off-diagonal vibronic coupling is taken to be linear in the vibrational coordinates,
\begin{equation}
\label{eq:Vab}
V_{ab} = \lambda_t^{ab} Q_t + \lambda_c^{ab} Q_c,
\end{equation}
where $\lambda_t^{ab}$ and $\lambda_c^{ab}$ are the vibronic coupling strengths between electronic states $a$ and $b$ through the two vibrational modes. In the present model, only selected pairs of electronic states are coupled, while all other off-diagonal couplings are set to zero. The detailed electronic energies, state-dependent vibronic couplings, and transition dipole matrix elements are listed in the SI. The bath Hamiltonian is modeled as a collection of harmonic oscillators,
\begin{equation}
\label{eq:HB}
H_B = \sum_q \sum_{\xi} \omega_{q\xi} b_{q\xi}^{\dagger}b_{q\xi},
\end{equation}
where $b_{q\xi}^{\dagger}$ and $b_{q\xi}$ are the creation and annihilation operators of the $\xi$-th bath mode associated with the $q$-th bath channel. The system--bath interaction is assumed to be linear in the bath coordinates,
\begin{equation}
\label{eq:HSB}
H_{SB} = \sum_q V_q \otimes B_q,
\qquad
B_q = \sum_{\xi} g_{q\xi} \left( b_{q\xi}^{\dagger} + b_{q\xi} \right).
\end{equation}
Here, $V_q$ are system operators chosen to be diagonal in the electronic basis and to act as identity operators in the explicit vibrational subspace. This form describes environmental fluctuations of electronic energy gaps and dephasing effects. The bath spectral density is described by the Drude form,
\begin{equation}
\label{eq:Drude}
J_q(\omega) = \frac{2\lambda_q\gamma_q\omega}{\omega^2+\gamma_q^2}.
\end{equation}
In the present calculations, three independent bath channels are used, with $\lambda_q=500~\mathrm{cm}^{-1}$ and $\gamma_q=1000~\mathrm{cm}^{-1}$ for all channels. This vibronic open-system Hamiltonian is then propagated using the hierarchical equations of motion to describe the combined effects of electronic coherence, vibronic coupling, relaxation, and environmental dephasing.

%
\begin{addendum} 
\item We thank Prof. Dr. Meng Zhou provide us the Au$_{25}$ complex. We also thank Prof. Dr. Wenwu Xu for the initial assessment of {ab-initio} calculations of Au complex. This work was supported by National Key Research and Development Program of China (Grant No.\ 2024YFA1409800),  NSFC Grant with No.\ 12274247, 12504279, Yongjiang talents program with No.\ 2022A-094-G, Ningbo International Science and Technology Cooperation with No.\ 2023H009 and the foundation of national excellent young scientist. The Next Generation Chemistry theme at the Rosalind Franklin Institute is supported by the EPSRC (V011359/1 (P)) (AJ). 

\item[Supporting information] The details of decay-time tables, wavelet analysis and oscillator spectra are shown in the SI. In addition, the data analysis of anti-diagonal bandwidth of main peaks, the transmission electron microscope and the TA before and after measuring are also revealed in the SI. 

\item[Competing Interests] The authors declare that they have no competing financial interests. 

\item[Correspondence] Correspondence of paper should be addressed to H.-G.D. ~(email: duanhongguang@nbu.edu.cn), F.Z. ~(zhengfulu@nbu.edu.cn) and A. J. ~(Ajay.Jha@rfi.ac.uk). 

\end{addendum}
%
\newpage

\newpage
%
\newpage
%
\newpage

\newpage

\end{document}